\documentclass{jfm}
\usepackage{color}
\usepackage{amstext}
\usepackage{amssymb}
\usepackage{graphicx}
\usepackage{amsmath}

\newcommand\const{\mathrm{const}}
\newcommand\Div{\mathrm{div}\,}

\newcommand\vV{\boldsymbol{V}}
\newcommand\va{\boldsymbol{a}}
\newcommand\vb{\boldsymbol{b}}
\newcommand\vc{\boldsymbol{c}}

\newcommand\vh{\boldsymbol{h}}

\newcommand\vj{\boldsymbol{j}}

\newcommand\vu{\boldsymbol{u}}

\newcommand\vx{\boldsymbol{x}}

\newcommand\vp{\boldsymbol{p}}
\newcommand\vq{\boldsymbol{q}}

\newcommand\vg{\boldsymbol{g}}

\newcommand\vomega{\boldsymbol{\omega}}

\newcommand\vxi{\boldsymbol{\xi}}

\begin{document}

{\title[MHD-Drift Equations: from Langmuir circulations to MHD-dynamo? ] {MHD-Drift Equations: from Langmuir
circulations to MHD-dynamo?}}

\author[V. A. Vladimirov]
{V.\ns A.\ns V\ls l\ls a\ls d\ls i\ls m\ls i\ls r\ls o\ls v}

\affiliation{Dept of Mathematics, University of York, Heslington, York, YO10 5DD, UK}

\pubyear{2010} \volume{xx} \pagerange{xx-xx}
\date{Sept 12th 2011}

\setcounter{page}{1}\maketitle \thispagestyle{empty}

\begin{abstract}

We have derived the closed system of averaged MHD-equations for general oscillating flows, which are purely
oscillating in the main approximation. We have used the mathematical approach which combines the two-timing
method and the notion of the distinguished limit. Properties of the commutators are used to simplify
calculations. The direct connection with a vortex dynamo (or the Langmuir circulations) has been demonstrated
and a conjecture on the MHD-dynamo has been formulated.

\end{abstract}

\section{Introduction \label{sect01}}

This paper derives the averaged MHD-equations for oscillating flows. The resulting equations are similar to
the original MHD-equations, but surprisingly (instead of commonly expected Reynolds stresses) the
\emph{drift velocity} (or just \emph{the drift}) plays a part of an additional advection velocity.

It is known that the drift can appear from either Lagrangian or Eulerian considerations. The
\emph{Lagrangian drift} appears as the average motion of Lagrangian particles and its theory is often based
on the averaging of ODEs, see \cite{Stokes, Lamb, LH, Batchelor, McIntyre, Craik, Yudovich}. In this paper we
focus on the \emph{Eulerian drift}, which  appears as the results of the Eulerian averaging of related PDEs
without addressing the motion of particles, see \cite{CraikLeib, Craik0, Riley, Vladimirov, Ilin}. The
detailed materials about the Eulerian drift can be found in \cite{Vladimirov}.

To  derive the averaged equations we employ the two-timing method, see \emph{e.g.}
\cite{Nayfeh, Kevor}.  We expose it as an elementary, systematic, and
justifiable procedure that follows the form developed by \cite{Yudovich, Vladimirov0,  Vladimirov1,
Vladimirov}. This mathematical procedure is complemented by a novel material on the distinguished limit,
which allows to find the proper slow time-scale.

\section{Functions and operations}

We introduce functions of variables $\vx=(x_1,x_2,x_3)$, $s$, and $\tau$, which in the text below serve as
dimensionless cartesian coordinates, slow time, and fast time.

\underline{\emph{Definition 1.}} The class $\mathbb{H}$ of \emph{hat-functions}  is defined
as
\begin{eqnarray}
\widehat{f}\in \mathbb{H}:\quad
\widehat{f}(\vx, s, \tau)=\widehat{f}(\vx,s,\tau+2\pi)\label{tilde-func-def}
\end{eqnarray}
where the $\tau$-dependence is always $2\pi$-periodic; the dependencies on $\vx$ and $t$ are not specified.

\underline{\emph{Definition 2.}}
For an arbitrary $\widehat{f}\in \mathbb{H}$ the \emph{averaging operation} is
\begin{eqnarray}
\langle {\widehat{f}}\,\rangle \equiv \frac{1}{2\pi}\int_{\tau_0}^{\tau_0+2\pi}
\widehat{f}(\vx, s, \tau)\,d\tau,\qquad\forall\ \tau_0\label{oper-1}
\end{eqnarray}
where  during the $\tau$-integration $s=\const$ and $\langle {\widehat{f}}\rangle$ does not depend on
$\tau_0$.

\underline{\emph{Definition 3.}} The class $\mathbb{T}$ of \emph{tilde-functions} is such that
\begin{eqnarray}
\widetilde f\in \mathbb{T}:\quad
\widetilde f(\vx, s, \tau)=\widetilde f(\vx,s,\tau+2\pi),\quad\text{with}\quad
\langle \widetilde f \rangle =0,\label{oper-2}
\end{eqnarray}
The tilde-functions are also called purely oscillating functions (\emph{$\mathbb{T}$-function} represents a
special case of $\mathbb{H}$-function with zero average).

\underline{\emph{Definition 4.}} The class $\mathbb{B}$ of \emph{bar-functions} is
defined as
\begin{eqnarray}
\overline{f}\in \mathbb{B}:\quad  \overline{f}_{\tau}\equiv 0,\quad
\overline{f}(\vx, s)=\langle\overline f(\vx,s)\rangle
 \label{oper-3}
\end{eqnarray}
(any $\mathbb{H}$-function can be uniquely separated into its $\mathbb{B}$- and $\mathbb{T}$- parts with the
use of (\ref{oper-1}))

\underline{\emph{Definition 5.}}
\emph{$\mathbb{T}$-integration (or tilde-integration)}: for a given $\widetilde{f}$ we introduce a new function
$\widetilde{f}^{\tau}$ called the $\mathbb{T}$-integral of $\widetilde{f}$:
\begin{eqnarray}
&&\widetilde{f}^{\tau}\equiv\int_0^\tau \widetilde{f}(\vx,s,\sigma)\,d\sigma
-\frac{1}{2\pi}\int_0^{2\pi}\Bigl(\int_0^\mu
\widetilde{f}(\vx,s,\sigma)\,d\sigma\Bigr)\,d\mu\label{oper-7}
\end{eqnarray}
which represents the unique solution of  a PDE $\partial
\widetilde{f}^{\tau}/\partial\tau=\widetilde{f}$ (with an unknown function $\widetilde{f}^{\tau}$
and a known function $\widetilde{f}$) supplemented by the condition $\langle \widetilde f\, \rangle=\langle
\widetilde f^\tau \rangle =0$ (\ref{oper-2}).

The $\tau$-derivative of $\mathbb{T}$-function always represents $\mathbb{T}$-function. However the
$\tau$-integration of $\mathbb{T}$-function can produce an $\mathbb{H}$-function. An example:
$\widetilde{f}=\overline{f}_1\sin\tau$ where $\overline{f}_1$ is an arbitrary $\mathbb{B}$: one can see that
$\langle\widetilde{f}\,\rangle\equiv 0$, however
$\langle\int_0^\tau\widetilde{f}(\vx,s,\sigma)d\sigma\rangle=\overline{f}_1\neq 0$, unless
$\overline{f}_1\equiv 0$. Formula (\ref{oper-7}) keeps the result of integration inside the
$\mathbb{T}$-class. $\mathbb{T}$-integration is inverse to $\tau$-differentiation
$(\widetilde{f}^{\tau})_{\tau}=(\widetilde{f}_{\tau})^{\tau}=\widetilde{f}$; the proof is omitted.

\underline{\emph{Definition 6.}}
A dimensionless function $f=f(\vx,s,\tau)$ belongs to the class $\mathbb{O}(1)$
\begin{eqnarray}
f\in \mathbb{O}(1)\label{all-one}
\end{eqnarray}
if $f={O}(1)$ and all required partial $\vx$-, $s$-, and $\tau$-derivatives of $f$ are also ${O}(1)$.

Here we emphasize that through all the text below all large or small parameters are represented by various
degrees of $\sigma$ only; these parameters appear as explicit multipliers in all formulae containing tilde-
and bar-functions; these functions always belong to $\mathbb{O}(1)$-class.

We also will use some properties of $\tau$-derivatives such as
\begin{eqnarray}
\widehat{f}_\tau=\widetilde{f}_\tau, \quad
\langle\widehat{f}_\tau\rangle=\langle\widetilde{f}_\tau\rangle=0 \label{oper-6}
\end{eqnarray}
The product of two $\mathbb{T}$-functions $\widetilde{f}$ and $\widetilde{g}$ represents a
$\mathbb{H}$-function: $\widetilde{f}\widetilde{g}\equiv\widehat{F}$, say. Separating $\mathbb{T}$-part
$\widetilde{F}$ from $\widehat{F}$ we write
\begin{eqnarray}
&&\widetilde{F}=\widehat{F}-\langle\widehat{F}\rangle
=\widetilde{f}\widetilde{g}-\langle\widetilde{f}\widetilde{g}\rangle=\{\widetilde{f}\widetilde{g}\}
\label{oper-5}
\end{eqnarray}
where the notation $\{\cdot\}$ for the tilde-part is introduced to avoid two levels of tildes. We will use
that the unique solution of a PDE inside the tilde-class is
\begin{eqnarray}
\partial \widetilde{f}/\partial\tau=0\quad \Rightarrow\quad \widetilde{f}\equiv 0\label{f-tilde=0}
\end{eqnarray}
which follows from (\ref{oper-7}). Since the average operation (\ref{oper-1}) is proportional to the
integration over $\tau$, then from the integration by parts we have
\begin{eqnarray}
&&\langle[\widetilde{\va},\widetilde{\vb}_\tau]\rangle=-\langle[\widetilde{\va}_\tau,\widetilde{\vb}]\rangle=-
\langle[\widetilde{\va}_\tau, \widehat{\vb}]\rangle,\
\langle[\widetilde{\va},\widetilde{\vb}^\tau]\rangle=-\langle[\widetilde{\va}^\tau,\widetilde{\vb}]\rangle=-
\langle[\widetilde{\va}^\tau, \widehat{\vb}]\rangle
\label{oper-15}
\end{eqnarray}
where $[\va,\vb]$ stands  for the commutator of two vector fields $\va$ and $\vb$ which is antisymmetric and
satisfies Jacobi's identity for vector fields $\va$, $\vb$, and $\vc$:
\begin{eqnarray}
&& [\va,\vb]\equiv(\vb\cdot\nabla)\va-(\va\cdot\nabla)\vb,\label{commutr}\\
&&[\va,\vb]=-[\vb,\va],\quad [\va,[\vb,\vc]]+ [\vc,[\va,\vb]]+[\vb,[\vc,\va]]=0 \label{oper-13}
\end{eqnarray}
A useful property of the commutator is
\begin{eqnarray}
&& \Div\va=0,\quad \Div\vb=0\quad \Rightarrow\quad \Div[\va,\vb]=0\label{oper-11a}
\end{eqnarray}
For any tilde-function $\widetilde{\va}$ and bar-function $\overline{\vb}$ (\ref{oper-15}),(\ref{oper-13})
give
\begin{eqnarray}
\langle[\widetilde{\va},[\overline{\vb},\widetilde{\va}^\tau]]\rangle=[\overline{\vb},\overline{\vV}_a]\quad
\text{where}\quad\overline{\vV}_a\equiv\frac{1}{2}\langle[\widetilde{\va},\widetilde{\va}^\tau]\rangle\label{lem1}
\end{eqnarray}

\section{Two-timing problem and distinguished limits}

The governing equation for MHD-dynamics of a homogeneous inviscid incompressible fluid with velocity field
$\vu^*$, magnetic fields $\vh^*$, vorticity $\vomega^*\equiv\nabla^*\times\vu^*$ and current
$\vj^*\equiv\nabla^*\times\vh^*$ is taken in the vorticity form
\begin{eqnarray}\label{exact-1}
&&\frac{\partial\vomega^*}{\partial {t}^*}+[\vomega^*,\vu^*]^*-[\vj^*,\vh^*]^* =0,\quad
\text{in}\quad \mathcal{D}^*\\
&&\frac{\partial\vh^*}{\partial {t}^*}+[\vh^*,\vu^*]^*=0,\quad\nonumber\\
&& \nabla^*\cdot\vu^*=0,\quad
\nabla^*\cdot\vh^*=0\nonumber
\end{eqnarray}
where asterisks mark dimensional variables, ${t}^*$-time, $\vx^*=(x_1^*,x_2^*,x_3^*)$-cartesian coordinates,
$\nabla^*=(\partial/\partial x_1^*, \partial/\partial x_2^*,\partial/\partial x_3^*)$,  and $[\,\cdot\, ,
\cdot\,]^*$ stands for the dimensional commutator (\ref{commutr}). In this paper we deal with
the transformations of equations, so the form of flow domain $\mathcal{D}^*$ and particular boundary
conditions can be specified at the later stages.

We accept that the considered class of (unknown) oscillatory solutions $\vu^*,\vh^*$ possesses characteristic
scales of velocity $U$, magnetic field $H$, length $L$, and high frequency $\sigma^*$
\begin{eqnarray}
&& U,\quad H,\quad L,\quad  \sigma^*\gg 1/T;\quad T\equiv L/U
\label{scales-list}
\end{eqnarray}
where $T$ is a dependent time-scale. In the chosen system of units the dimensions of $U$ and $H$ coincide; we
choose them being the same order $U=H$. The dimensionless variables and frequency are
\begin{eqnarray}
&& \vx\equiv\vx^*/L,\quad t\equiv t/T,
\quad\widehat{\vu}\equiv\widehat{\vu}^*/U,\quad\widehat{\vh}\equiv\widehat{\vh}^*/U,\quad\sigma\equiv\sigma^*T\gg
1
\label{scales}
\end{eqnarray}
We assume that the flow has its own intrinsic slow-time scale $T_{\text{slow}}$ (which can be different from
$T$) and consider solutions of (\ref{exact-1}) in the form of hat-functions (\ref{tilde-func-def})
\begin{eqnarray}
&& \vu^*=U \widehat{\vu}(\vx, s, \tau),\quad \vh^*=U \widehat{\vh}(\vx, s, \tau);\quad  \text{with}\, \
\tau\equiv\sigma {t},\, s\equiv\Omega\, {t},
\, \Omega\equiv T/T_{\text{slow}}\label{exact-2}
\end{eqnarray}
Then the use of the chain rule and transformation to dimensionless variables give
\begin{eqnarray}
&&\left(\frac{\partial}{\partial\tau}+
\frac{\Omega}{\sigma}\frac{\partial}{\partial s}\right)\widehat{\vomega}+
\frac{1}{\sigma}[\widehat{\vomega},\widehat{\vu}]-\frac{1}{\sigma}[\widehat{\vj},\widehat{\vh}]
= 0\label{exact-6-h}\\
&&\left(\frac{\partial}{\partial\tau}+
\frac{\Omega}{\sigma}\frac{\partial}{\partial s}\right)\widehat{\vh}+
\frac{1}{\sigma}[\widehat{\vh},\widehat{\vu}]= 0\nonumber
\\
&&\Div\widehat{\vh}=0,\quad \Div\widehat{\vu}=0\nonumber
\end{eqnarray}
In order to keep variable $s$ `slow' in comparison with $\tau$ we have to accept that $\Omega/\sigma\ll 1$.
Then eqn.(\ref{exact-6-h}) contains two independent small parameters:
\begin{eqnarray}
\varepsilon\equiv \frac{1}{T\sigma^*}=\frac{1}{\sigma},\quad \varepsilon_1\equiv
\frac{1}{T_\text{slow}\sigma^*}\equiv\frac{\Omega}{\sigma}\label{small-pars-h}
\end{eqnarray}
Here we must make an auxiliary (technically essential) assumption: after the use of the chain rule
(\ref{exact-6-h}) variables $s$ and $\tau$  are (temporarily) considered to be \emph{mutually independent}:
\begin{eqnarray}
&& \tau,\qquad s \qquad -\text{independent variables}\label{tau-s-ind}
\end{eqnarray}
From the mathematical viewpoint the increasing of the number of independent variables in a PDE represents a
very radical step, which leads to an entirely new PDE. This step should be justified \emph{a posteriori} by
the estimations of the error of the obtained solution (rewritten back to the original variable $t$)
substituted to the original equation (\ref{exact-1}).

In a rigorous asymptotic procedure with $\sigma\to\infty$ one has to consider asymptotic paths on the
$(\varepsilon, \varepsilon_1)$-plane such that
\begin{eqnarray}
&& (\varepsilon, \varepsilon_1)\to (0,0)\label{paths}
\end{eqnarray}
Each such path can be prescribed by a particular function $\Omega(\sigma)$. One may expect that there are
infinitely many different (although some of them can coincide) solutions to (\ref{exact-6-h}) corresponding
to different $\Omega(\sigma)$. However for these equations (as well as for many others)  a unique path can be
found, which is called the \emph{distinguish limit} (or the \emph{distinguished path}).  The notion of the
\emph{distinguish limit} is practical and heuristic, see \cite{Nayfeh, Kevor}; its definition can vary for
different equations and in different books and papers. For our problem we write that \emph{the distinguished
limit  is given by such a function $\Omega=\Omega_d(\sigma)$ that allows to build a self-consistent
asymptotic solution}. Here the term
\emph{self-consistent asymptotic solution} means that the required successive approximations can be calculated.
These calculations  include the elimination of the reducible secular in $s$ terms; the
\emph{reducible secular terms} are such terms which can be excluded by increasing the slow time-scale. For
instance, a non-secular term proportional to $\sin s$ gives secular terms in its Taylor's decomposition with
respect to $t=\sigma s$). Below we show that the choice
\begin{eqnarray}
\Omega(\sigma)=1/\sigma:\qquad\tau=\sigma t,\quad s=t/\sigma,\qquad \alpha=\const\label{tau-s-h}
\end{eqnarray}
allows to build the distinguished limit solution. The uniqueness of such a path for the considered class of
solutions can be proven but we avoid such details in this paper. Hence the governing equations are
\begin{eqnarray}
&&\widehat{\vomega}_\tau+
\varepsilon[\widehat{\vomega},\widehat{\vu}]-\varepsilon[\widehat{\vj},\widehat{\vh}]
+\varepsilon^2\widehat{\vomega}_s= 0,
\quad \varepsilon\equiv 1/\sigma\to 0\label{exact-6eps}\\
&&\widehat{\vh}_\tau+
\varepsilon[\widehat{\vh},\widehat{\vu}]+\varepsilon^2\widehat{\vh}_s= 0\nonumber\label{exact-6eps-h}\\
&&\Div\widehat{\vu}=0, \quad \Div\widehat{\vh}=0\nonumber
\end{eqnarray}
where the subscripts $\tau$ and $s$ denote the related partial derivatives.

\section{Derivation of the MHD-Drift averaged equation \label{sect04}}

Let us look for solutions of (\ref{exact-6eps}) in the form of regular series
\begin{eqnarray}
&&(\widehat{\vh},\widehat{\vu})=\sum_{k=0}^\infty\varepsilon^k (\widehat{\vh}_k,\widehat{\vu}_k);\quad
\widehat{\vh}_k, \widehat{\vu}_k\in \mathbb{H}\cap\mathbb{O}(1),\quad k=0,1,2,\dots
\label{basic-4aa-0-h}
\end{eqnarray}
In this paper we enforce the restriction
\begin{eqnarray}
&& \overline{\vu}_0\equiv 0,\quad \overline{\vh}_0\equiv 0\label{mean0}
\end{eqnarray}
which is natural physically if one considers, say, how the secondary vorticity develops on the background of
a wave motion. The substitution of (\ref{basic-4aa-0-h}),(\ref{mean0}) into (\ref{exact-6eps}) produces the
equations of successive approximations. The equations of zero approximation are
\begin{eqnarray}
&&\widehat{\vomega}_{0\tau}=\widetilde{\vomega}_{0\tau}=0;\quad
\widehat{\vh}_{0\tau}=\widetilde{\vh}_{0\tau}=0\label{eqn-0-0-h}
\end{eqnarray}
Their unique solution (\ref{f-tilde=0}) is $\widetilde{\vomega}_{0}\equiv 0$ and $\widetilde{\vh}_{0}\equiv
0$. Taking into account (\ref{mean0}) we can write
\begin{eqnarray}
&&\widehat{\vomega}_0\equiv 0, \quad \widehat{\vh}_0\equiv 0\label{sol-0}
\end{eqnarray}
which means that in zero approximation the flow  is potential, purely oscillating, and the magnetic field
vanishes. This leads to the similar equations for the first approximation of
(\ref{exact-6eps}),(\ref{basic-4aa-0-h})-(\ref{sol-0})
\begin{eqnarray}
&&\widehat{\vomega}_{1\tau}=0;\quad\widehat{\vh}_{1\tau}=0\label{eqn-0-1-h}
\end{eqnarray}
which have the unique solution
\begin{eqnarray}
&&\widetilde{\vomega}_1\equiv 0,\quad
\widetilde{\vh}_1\equiv 0,\quad\overline{\vomega}_1=\boxed{?},\quad\overline{\vh}_1=\boxed{?}\label{sol-1}
\end{eqnarray}
where  mean functions remain undetermined. The equations of second approximation that take into account
(\ref{mean0}),(\ref{sol-0}),(\ref{sol-1}) are
\begin{eqnarray}
&&\widetilde{\vomega}_{2\tau}+[\overline{\vomega}_1,\widetilde{\vu}_0]=0,\quad
\widetilde{\vomega}_{2\tau}+[\overline{\vomega}_1,\widetilde{\vu}_0]=0
\label{eqn-2}
\end{eqnarray}
which after  $\mathbb{T}$-integration (\ref{oper-7}) yield
\begin{eqnarray}
&&\widetilde{\vomega}_2=[\widetilde{\vu}_0^\tau,\overline{\vomega}_1],\quad
\widetilde{\vh}_2=[\widetilde{\vu}_0^\tau,\overline{\vh}_1],\quad
\overline{\vomega}_2=\boxed{?},\quad\overline{\vh}_2=\boxed{?}\label{sol-2}
\end{eqnarray}
The equations of third approximation that take into account (\ref{mean0}),(\ref{sol-0}),(\ref{sol-1}) are
\begin{eqnarray}
&&\widetilde{\vomega}_{3\tau}+\overline{\vomega}_{1s}+[\widehat{\vomega}_2,\widetilde{\vu}_0]+
[\overline{\vomega}_1,\widehat{\vu}_1]-[\overline{\vj}_1,\overline{\vh}_1]=0
\label{eqn-3}\\
&&\widetilde{\vh}_{3\tau}+\overline{\vh}_{1s}+[\widehat{\vh}_2,\widetilde{\vu}_0]+
[\overline{\vh}_1,\widehat{\vu}_1]=0\nonumber
\label{eqn-3}
\end{eqnarray}
The bar-part (\ref{oper-1}) of this system is
\begin{eqnarray}
&&\overline{\vomega}_{1s}+
[\overline{\vomega}_1,\overline{\vu}_1]-[\overline{\vj}_1,\overline{\vh}_1]+
\langle[\widetilde{\vomega}_2,\widetilde{\vu}_0]\rangle=0\label{eqn-3-bar}\\
&&\overline{\vh}_{1s}+ [\overline{\vh}_1,\overline{\vu}_1]+
\langle[\widetilde{\vh}_2,\widetilde{\vu}_0]\rangle=0\nonumber
\label{eqn-3-tilde}
\end{eqnarray}
which can be transformed with the use of (\ref{sol-2}) and (\ref{lem1}) to the final form
\begin{eqnarray}
&&\overline{\vomega}_{1s}+
[\overline{\vomega}_1,\overline{\vu}_1+\overline{\vV}_0]-[\overline{\vj}_1,\overline{\vh}_1]=0
\label{eqn-3-bar1-1}\\
&&\overline{\vh}_{1s}+ [\overline{\vh}_1,\overline{\vu}_1+\overline{\vV}_0]=0
\nonumber\\
&&\overline{\vV}_0\equiv\frac{1}{2}\langle[\widetilde{\vu}_0,\widetilde{\vu}_0^\tau]\rangle\label{drift-vel}
\end{eqnarray}
If one uses these equations as a closed mathematical model, then all the subscripts and bars can be deleted:
\begin{eqnarray}
&&\vomega_{s}+ [\vomega,\vu+\vV]-[\vj,\vh]=0
\label{eqn-3-bar1-1-Model}\\
&&\vh_{s}+ [\vh,\vu+\vV]=0
\nonumber\\
&&\vV\equiv\frac{1}{2}\langle[\widetilde{\vu},\widetilde{\vu}^\tau]\rangle\label{drift-vel-Model}
\end{eqnarray}
One can see that:

1. The equation for the oscillating velocity $\widetilde{\vu}_0$ in our consideration is absent, there are
only two restrictions:  $\widetilde{\vu}_0$ is incompressible and potential. Hence the drift velocity
$\overline{\vV}_0$ (\ref{drift-vel}) represents a function that is `external' to the equations.

2. The derived system of equations (\ref{eqn-3-bar1-1}) looks similar to the original one (\ref{exact-1}).
One may think that (\ref{eqn-3-bar1-1}) describes `just' an additional advection of vorticity and magnetic
field by the drift. However, the fact that the averaged vorticity is transported with such an additional
velocity is highly non-trivial; in particular, it contains the possibility of the \emph{vortex dynamo} or
Langmuir circulations, which we consider below.

\section{Stokes drift}

Our description of the drift differs from a classical one, therefore we first demonstrate the match of
(\ref{drift-vel}) with the classical Stokes drift. Let velocity field $\widetilde{\vu}_0$ and
$\widetilde{\vxi}_0\equiv\widetilde{\vu}^\tau_0$ be
\begin{eqnarray}
&&\widetilde{\vu}_0(\vx,s,\tau)=\overline{\vp}(\vx,t)\sin\tau+
\overline{\vq}(\vx,t)\cos\tau  \label{Example-1-1}\\
&&\widetilde{\vxi}_0(\vx,t,\tau)=-\overline{\vp}(\vx,t)\cos\tau+
\overline{\vq}(\vx,t)\sin\tau\nonumber\label{Example-1-2}
\end{eqnarray}
with arbitrary $\mathbb{B}$-functions  $\overline{\vp}$ and $\overline{\vq}$. Straightforward calculations yield
\begin{eqnarray}
&[\widetilde{\vu}_0,\widetilde{\vxi}_0]=[\overline{\vp},\overline{\vq}]\label{Example-1-3}
\end{eqnarray}
hence  the commutator is surprisingly not oscillating. The drift velocity (\ref{drift-vel}) is
\begin{eqnarray}
&&\overline{\vV}_0=\frac{1}{2}\langle[\widetilde{\vu}_0,\widetilde{\vxi}_0]\rangle=
\frac{1}{2}[\overline{\vp},\overline{\vq}]\label{Example-1-5}
\end{eqnarray}
The \emph{dimensional} solution for a  plane potential harmonic travelling wave is
\begin{equation}\label{Stokes}
   \widehat{\vu}^*_0=U\widetilde{\vu}_0,\quad ,
   \quad\widetilde{\vu}_0=\exp(k^*z^*)\left(\begin{array} {c} \cos(k^*x^*-\tau)\\ \sin (k^*x^*-\tau)\end{array}\right)
\end{equation}
where $(x^*,z^*)$ are cartesian coordinates and $k^*=1/L$ is a wavenumber. In \cite{Stokes, Lamb, Debnath}
one can see that $U={k^*g^*a^*}/{\sigma^*}$ where $a^*$ and $g^*$ are dimensional spatial wave amplitude and
gravity; however these physical details are excessive for our analysis. The dimensionless velocity field
(\ref{Stokes}) and $\widetilde{\vxi}$ are
\begin{eqnarray}
\widetilde{\vu}_0=e^{z}\left(\begin{array} {c} \cos(x-\tau)\\ \sin (x-\tau)\end{array}\right),\quad
\widetilde{\vxi}_0=e^{z}\left(\begin{array}{c} -\sin(x-\tau)\\ ~\cos (x-\tau)\end{array}\right)
\label{Example-3-1}
\end{eqnarray}
where both fields (\ref{Example-3-1}) are unbounded as $z\to\infty$, but it is not essential for our
purposes. Fields $\overline{\vp}(x,z)$, $\overline{\vq}(x,z)$ (\ref{Example-1-1}) are
\begin{eqnarray}
\overline{\vp}=Ae^{z}\left(\begin{array}{c} \sin x\\ -\cos x\end{array}\right),\quad
\overline{\vq}=Ae^{z}\left(\begin{array}{c} \cos x\\ \sin x\end{array}\right)
\label{Example-3-2}
\end{eqnarray}
The calculations (with the use of (\ref{Example-1-3})) yield
\begin{eqnarray}
\overline{{\vV}}_0= e^{2z}\left(\begin{array}{c} 1\\ 0\end{array}\right)
\label{Example-3-3}
\end{eqnarray}
The dimensional version of (\ref{Example-3-3}) is
\begin{eqnarray}
\overline{{\vV}}_0^*= \frac{U^2k^*}{\sigma^*}e^{2k^*z^*}\left(\begin{array}{c} 1\\ 0\end{array}\right)
\label{Example-3-3d}
\end{eqnarray}
which coincides with the classical expression for the drift velocity given by \cite{Stokes, Lamb, Debnath}.
To obtain (\ref{Example-3-3d}) one should take into account the difference between the time $t$ and
$s=t/\sigma$ (\ref{tau-s-h}).

\section{ The averaged Euler's equations  and vortex dynamo}

A special case of (\ref{eqn-3-bar1-1}),(\ref{eqn-3-bar1-1-Model}) without a magnetic field is
\begin{eqnarray}
&&\vomega_{s}+ [\vomega,\vu+\vV]=0.\quad\Div\vu=0
\label{eqn-3-bar1-1-E}
\end{eqnarray}
Different $s$-independent versions of eqn.(\ref{eqn-3-bar1-1}) were derived  in the studies of Langmuir
circulation by \cite{CraikLeib} and for the steady streaming problems by
\cite{Riley, Ilin}; the methods employed by these authors are different and more cumbersome than our method.
In order to demonstrate the possibility of vortex dynamo we first notice that eqn.(\ref{eqn-3-bar1-1-E}) can
be integrated (in space) as
\begin{eqnarray}
&&\vu_{s}+ (\vu\cdot\nabla)\vu+\vomega\times\vV=-\nabla p,\quad \nabla\cdot\vu=0
\label{eqn-3-bar1-vel}
\end{eqnarray}
where $\overline{p}$ is a function of integration and the second equation follows from the continuity
equation in (\ref{exact-1}). Let the zero approximation (\ref{sol-0}) represent the plane potential
travelling gravity wave (\ref{Example-3-1}) with the drift velocity (\ref{Example-3-3}). Let cartesian
coordinates $(x,y,z)$ be such that $\vV=(U, 0,0)$, $U=e^{2z}$, $\vu=(u,v,w)$ where all components are
$x$-independent (translationally-invariant), and $x,z$-variables coincide with ones in (\ref{Example-3-1}).
Then the component form of (\ref{eqn-3-bar1-vel}) is
\begin{eqnarray}
&&u_s+vu_y+wu_z=0\nonumber\\
&&v_s+uv_y+wv_z-Uu_y=-\overline{p}_y\nonumber\\
&&w_s+vw_y+ww_z-Uu_z=-\overline{p}_z\nonumber \\
&&v_y+w_z=0\nonumber
\end{eqnarray}
which can be rewritten as (see \cite{VladimirovS, VladimirovS1})
\begin{eqnarray}
&&v_s+vv_y+wv_z=-P_y-\rho \Phi_y\label{strat-eqns}\\
&&w_s+vw_y+ww_z=-P_z-\rho \Phi_z\nonumber\\
&&v_y+w_z=0\nonumber\\
&&\rho_s+u\rho_x+v\rho_y=0\nonumber
\end{eqnarray}
where $\rho\equiv u$, $\Phi\equiv U=e^{2z}$, and $P$ is a modified pressure. One can see that
(\ref{strat-eqns}) is mathematically equivalent to the system of equations for an incompressible stratified
fluid, written in Boussinesq's approximation. The effective `gravity field' $\vg=-\nabla\Phi=(0, 0,
-2e^{2z})$ is non-homogeneous that makes the analogy with a `standard' stratified fluid non-complete.
Nevertheless one can see that any increasing function $u(z)\equiv\rho(z)$ (taken from the shear flow
$(u,v,w)=(u(z),0,0)$) produces `Taylor instability' of an inversely stratified equilibrium. It leads to the
growth of longitudinal vortices and can be connected to Langmuir circulations, see
\cite{CraikLeib, Leibovich, Craik0,  Thorpe}.

\section{Discussion}

1. The consideration of this paper is based on the assumption that the enforced frequency $\sigma^*$
(\ref{scales}) of oscillations is higher than all intrinsic frequencies. This frequency appears in our theory
via the prescribed potential velocity $\widetilde{\vu}_0$  (\ref{eqn-0-0-h}).

2. The prescribed oscillatory velocity $\widetilde{\vu}_0$ can be caused by different factors. For example,
it can be produced by oscillations of boundaries or appear in full viscous theory after the matching of
external flow with boundary-layer solution. The latter option is often considered, see
\cite{Riley,Vladimirov1,Ilin}.

3.  To  justify the distinguished limit (\ref{tau-s-h}) mathematically, one should prove that any different
path $\Omega(\sigma)$ produces an asymptotic solution which contains terms secular in $s$  or does not
produce any asymptotic solution at all. The following statement can be proven: for $\overline{\vV}_0\neq 0$
(\ref{drift-vel}) and the function $\Omega(\sigma)=1/\omega^\alpha$ (with a constant $\alpha>-1$) all
solutions with $\alpha<1$ contain secular terms, while all equations with $\alpha>1$ produce a controversial
(unsolvable) equations of successive approximations. If $\overline{\vV}_0\equiv 0$ then the statement is
different but we do not describe it here. The proof is omitted.

4. There is a challenging physical fact to explain: the existence of the distinguished limit (\ref{tau-s-h})
means that there is a hidden slow time-scale $T_{\text{slow}}=\sigma^* T^2$ in the system.

5. The consideration of translationally-invariant MHD-motion in (\ref{eqn-3-bar1-1}) is possible in the
spirit of the analogy between MHD flows and stratified flows, see \cite{VladMof}.

6. The mathematical justification of the equation (\ref{eqn-3-bar1-1}) by the estimation of the error in the
original equation (\ref{exact-1}) is easily achievable.

7. The higher approximations of the averaged  equation (\ref{eqn-3-bar1-1}) can be derived. They are
especially useful for the study of motions with $\overline{\vV}_0\equiv 0$. In particular, one can show that
in this case Langmuir circulations cab be still generated by a similar mechanism.

8. The viscosity and diffusivity can be routinely incorporated in (\ref{eqn-3-bar1-1}) as the RHS-terms
$\nu\nabla^2\overline{\vomega}_1$ and  $\kappa\nabla^2\overline{\vh}_1$  . Accordingly,  viscous and
diffusion terms will appear in the equations (\ref{eqn-3-bar1-vel}) and (\ref{strat-eqns}). At the same time
after the incorporation of viscosity one more small parameter appears in the list (\ref{small-pars-h}), and
the distinguished limit should be reconsidered.

9. The incorporation of the density stratification and gravity field into presented theory (or as a separate
theory) is straightforward.

10. The abolishing of the requirement of a vanishing mean flow in zeroth approximation (\ref{mean0}) is also
straightforward. However in this case the distinguished limit (\ref{tau-s-h}) is different and the resulted
averaged equations are more complicated than (\ref{eqn-3-bar1-1}).

11. For the finite and time-dependent flow domain $\mathcal{D}(t)$ the definition of average (\ref{oper-1})
directly works only if $\vx\in\mathcal{D}$ at any instant. If at some instant $\vx\notin\mathcal{D}$ then the
theory should include a `projection' of the boundary condition on the `undisturbed' boundary. Such a
consideration requires the smallness of the amplitude $a^*/L$ of spatial oscillations of fluid particles.
However one can see that $a^*\sim \vu^*/\sigma^*$ and hence $a^*/L\sim 1/\sigma\equiv \varepsilon$
(\ref{small-pars-h}). Therefore the considering of a time-dependent domain does not introduce any new small
parameter, and the distinguished limit (\ref{tau-s-h}) will stay the same. In particular, it means that our
small parameter $\varepsilon$ is the same as the dimensional slope of free surface in the theory of Langmuir
circulations by \cite{CraikLeib}.

12. The determining of the function $u(z)\equiv\rho(z)$ in (\ref{strat-eqns}) (for real Langmuir
circulations) requires an additional theory which includes viscosity or turbulent tangential stresses at the
free surface. It is interesting, that from this viewpoint an `unstable stratification' can be continuously
generated and amplified by tangential stresses applied at free surface. We do not compete here with the
theories by \cite{CraikLeib, Leibovich, Thorpe}, which describe this phenomenon  well.

13. One can suggest that since the equations (\ref{eqn-3-bar1-1}) for $\vh\equiv 0$   do describe a mechanism
of vortex dynamo, and the mathematical structure of the full averaged equations
(\ref{eqn-3-bar1-1}),(\ref{eqn-3-bar1-1-Model}) with $\vh\ne 0$ is similar, then these full equations could
also describe a possible mechanism of MHD-dynamo, such as the generation of the magnetic field of the Earth.

\begin{acknowledgments}
The author thanks the Department of Mathematics of the University of York for the research-stimulating
environment. The author is grateful to Profs. A.D.D.Craik, K.I.Ilin, S.Leibovich, H.K.Moffatt, A.B.Morgulis,
N.Riley, and V.A.Zheligovsky for helpful discussions.
\end{acknowledgments}

\end{document}